\documentclass{elsart}

\usepackage{epsfig}

\usepackage{amssymb,amsmath}

\usepackage{url}

\begin{document}

\begin{frontmatter}



\title{Probing photon helicity in radiative 
$\boldsymbol{B}$ decays via charmonium resonance interference}


\author{Mathias~Knecht,}
\author{Thomas~Schietinger}

\address{Laboratory for High-Energy Physics,
Ecole Polytechnique F\'ed\'erale,\\ CH-1015 Lausanne, Switzerland}

\begin{abstract}
We investigate a new method to probe the helicity of the photon emitted
in the $b\rightarrow s\gamma$ transition.
The method relies on the observation of interference effects between 
two resonance contributions, $B \rightarrow K^*(K\gamma)\gamma$ and 
$B \rightarrow \eta_c(\gamma\gamma)K$ or 
$B \rightarrow \chi_{c0}(\gamma\gamma)K$
to the same final state $K\gamma\gamma$. 
Decays of the type $B \rightarrow K_\text{res}(K\gamma)\gamma$ dominate the
$B\rightarrow K\gamma\gamma$ yield throughout most of the phase space, and may
be accessible at current $B$ meson facilities already.
\end{abstract}

\end{frontmatter}


\section{Introduction}

Flavor-changing neutral currents are an important testing ground for
the Standard Model (SM) of elementary particles.
The quark transition $b\rightarrow s \gamma$ has played an outstanding role in 
this respect by providing direct experimental evidence for the penguin
diagram process, which is expected to be particularly sensitive to 
contributions from physics beyond the SM.
Recent measurements of the $b\rightarrow s\gamma$ rate \cite{exp},
however, agree very well with theoretical predictions \cite{theor}, leaving
little hope for observing hints of new physics via the decay rate only.
Consequently, recent efforts have focused on finding additional 
observable degrees of freedom related to $b\rightarrow s\gamma$, such as CP 
asymmetries or the helicity of the emitted photon, in order to subject the
SM to ever more stringent tests.
In a similar vein, the decay $B\rightarrow X_s\gamma\gamma$ and its exclusive manifestation
$B\rightarrow K\gamma\gamma$ have been studied in this context \cite{RRS,Choudhury,Hiller-Safir}.
In analogy to $b\rightarrow sl^+l^-$, the diphoton invariant mass spectrum and
forward-backward asymmetries have been suggested as probes for new physics beyond 
the SM \cite{Cao-Xiao-Lu-Gemintern}.

In this Letter, we point out the significance of contributions to the $K\gamma\gamma$
final state that occur via radiatively decaying kaon resonances: 
$B\rightarrow K_\mathrm{res}\gamma$, with $K_\mathrm{res}$ being any kaon resonance, 
such as $K^*(892)$ or higher, that can decay to $K\gamma$.
We will further show how these decays may be used to extract information on 
the helicity of the emitted photon in the $b\rightarrow s\gamma$ amplitude
at future high-statistics $B$-meson facilities.

It was first noted by Atwood, Gronau, and Soni \cite{Atwood-Gronau-Soni}
that the photon helicity in $b\rightarrow s\gamma$
carries information on the underlying interaction.
While the SM amplitude for $b\rightarrow s\gamma$ results in a predominantly
left-handed photon (right-handed for $\overline{b}\rightarrow \overline{s}\gamma$), 
there are extensions of the SM that could alter the helicity of the photon 
without affecting much the rate of the decay.
Thus several methods for an indirect determination of the photon helicity
in radiative $B$ decays have been devised:
1) study of the \textit{interference between $b\rightarrow s\gamma$ and
$\overline{b}\rightarrow \overline{s}\gamma$}, made possible by the phenomenon of
$B^0$--$\overline{B}\mbox{}^0$ mixing \cite{Atwood-Gronau-Soni};
2) analysis of the \textit{decay photon by means of its conversion to $e^+e^-$}
\cite{Grossman-Pirjol}
(see also \cite{Melikhov-Nikitin-Simula} for the case of off-shell photons);
3) analysis of the \textit{recoil system arising from the hadronization
of the $s$-quark} in $b\rightarrow s\gamma$ \cite{GGPR};
4) use of a \textit{polarized initial state}, \emph{i.e.} $b$-baryon decay,
to infer the photon polarization from angular correlations with the final state
\cite{Mannel-Recksiegel,Hiller-Kagan}.
Yet another way to analyze the decay photon is provided by the interference
with another photon in a well-known state arising
from the \textit{same decay}.
For example $B\rightarrow K^*(K\gamma)\gamma$ can interfere with
$B\rightarrow K c\bar{c}(\gamma\gamma)$, where $c\bar{c}$ is a charmonium
state such as $\eta_c$ or $\chi_{c0}$.

Photon pairs arising from $\eta_c$ ($J^P = 0^-$) decay are 
known to be in an exact state of perpendicular polarization \cite{Yang}, 
\emph{i.e.} a state with photon spin orientation given by
$\mathbf{k_1\cdot[\boldsymbol{\epsilon}_1(k_1) \times \boldsymbol{\epsilon}_2(k_2)]}$, where  
$\mathbf{\boldsymbol{\epsilon}_1}$ and $\mathbf{\boldsymbol{\epsilon}_2}$ 
($\mathbf{k_1}$ and $\mathbf{k_2}$) are the transverse polarization (momentum) 
vectors of the two photons.
Similarly, photons from $\chi_{c0}$ ($J^P = 0^+$) decay are in a state of 
parallel polarization ($\mathbf{\boldsymbol{\epsilon}_1\cdot\boldsymbol{\epsilon}_2}$).
Thus we may use $\eta_c$ and $\chi_{c0}$ as probes to analyze the polarization
state of the photons from $B\rightarrow K^*(K\gamma)\gamma$,
since photons from $\eta_c$ ($\chi_{c0}$) will only interfere with the perpendicular 
(parallel) polarization component. 


\section{\boldmath{$B\rightarrow K^*(K\gamma)\gamma$} amplitude}

The SM amplitude for $B\rightarrow K^*(K\gamma)\gamma$ as given in 
Ref.~\cite{Choudhury} is based on a description of 
$b\rightarrow s\gamma$ in the framework of a leading-order effective Hamiltonian,
\begin{equation}
    \mathcal{H}_\mathrm{eff} = 
     -4 \frac{G_F}{\sqrt{2}} V_{tb} V_{ts}^* C_7 O_7 
\label{eq:Heff}
\end{equation} 
with $G_F$ the Fermi constant, $C_7$ the Wilson coefficient of the local operator 
$O_7= (e m_b)/(16\pi^2)\overline{s}_L \sigma_{\mu\nu} b_R F^{\mu\nu}$, 
$e$ the electric charge, $m_b$ the mass of the $b$-quark,
$F^{\mu\nu}$ the electromagnetic field tensor and $\sigma_{\mu\nu} = 
\frac{i}{2}(\gamma_\mu\gamma_\nu - \gamma_\nu\gamma_\mu)$.
$V_{tb}$ and $V_{ts}$ are the usual Cabibbo-Kobayashi-Maskawa matrix elements.
The full amplitude is then given as
 ${\mathcal M}_{K^*} = \left[T^{\mu\nu}(k_1,k_2) +  T^{\nu\mu}(k_2,k_1)\right]
                   \epsilon^*_\mu(k_1) \epsilon^*_\nu(k_2)$
with 
\begin{multline} 
    T^{\mu\nu}(k_1,k_2) =  
    \left(\frac{e m_b g F}{16 \pi^2}\right) 
    4\frac{G_F}{\sqrt{2}}V_{tb}V^*_{ts} C_7 
   \epsilon^{\alpha\nu\gamma\delta} 
    k_{2\alpha} (p_B-k_1)_\gamma k_{1\beta '} \\
    \times \frac{\left( g_{\delta \sigma '}-\frac{(p_B-k_1)_\delta(p_B-k_1)_{\sigma '}}{m^2_{K^*}}\right)}
       {(p_B-k_1)^2-m^2_{K^*}+im_{K^*}\Gamma_{K^*}} \\ 
    \times \left[i \epsilon^{\mu \beta'\sigma'\tau'} (p_B-k_1)_{\tau'}-  \right.  
       \left. \left(g^{\mu \sigma'}(p_B-k_1)^{\beta'}-g^{\beta'\sigma'}(p_B-k_1)^{\mu}\right)\right]
\label{eq:SMamp}
\end{multline}
where $k_i$ are the 4-vectors $(E,\mathbf{p})$ of the photons, and $p_B$, $p_K$ the 4-vectors of the $B$ and $K$
mesons. 
The constants $g$ and $F$ are related to the coupling strengths for $K^* \rightarrow K\gamma$ and 
$B \rightarrow K^*\gamma$, respectively, and are different for neutral ($B^0$) and charged decays
($B^+$).

\begin{figure}[t]
\center
\includegraphics*[width=12cm]{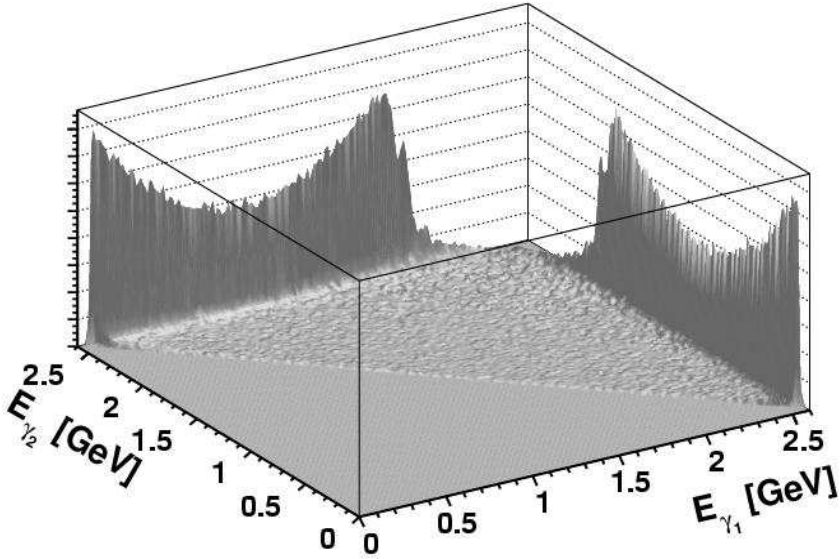}  
\caption{Decay distribution for $B\rightarrow K^*\gamma \rightarrow K \gamma\gamma$
 in the plane of the two photon energies (Dalitz plot).} 
\label{fig:SMdalitz}
\end{figure}

The decay distribution in the plane of the two photon energies (Dalitz plot)
is shown in Fig.~\ref{fig:SMdalitz}.
It exhibits the typical $(1+\cos^2\theta)$ shape along the resonance lines, as 
expected for the decay of a pseudoscalar particle into a pseudoscalar and two vectors
via an intermediate vector resonance state.
It also features a non-negligible fraction of decays in the central region
of the Dalitz plot, outside the two resonance lines. 
This region is populated by decays receiving contributions
from {\it both} amplitudes, 
$B \rightarrow K^* \gamma \rightarrow (K \gamma') \gamma$ and
$B \rightarrow K^* \gamma'\rightarrow (K \gamma)  \gamma'$.
Despite the suppression from the Breit-Wigner resonance shape the
effect of this interference amplitude results in a substantial
enhancement of the over-all branching fraction of the decay.
Indeed, from the distribution of events we find that
 $\mathcal{B}(B \rightarrow K^*(K\gamma)\gamma) \approx 
  3.85 \times
  \mathcal{B}(B \rightarrow K^*\gamma) \times
  \mathcal{B}(K^{*} \rightarrow K\gamma)$.
Combining this estimate with recent experimental data on 
$B\rightarrow K^*\gamma$ \cite{HFAG} and $K^*\rightarrow K\gamma$ \cite{PDG}
we obtain branching fractions of (3.54 $\pm$ 0.35) for $B^0$ and
(1.54 $\pm$ 0.15) for $B^+$ in units of 10$^{-7}$, well accessible with 
the next generation of $B$ factories \cite{SuperB} and perhaps also 
at hadron colliders \cite{LHCbReoptTDR} if backgrounds can be controlled.

\section{Other contributions to \boldmath{$B\rightarrow K\gamma\gamma$}}

Other transitions yielding the $K\gamma\gamma$ final state include 
a non-resonant (short-distance) contribution,
$b\rightarrow s\gamma$ contributions via higher kaon resonances decaying to $K\gamma$,
contributions from $\eta(\gamma\gamma)K$ and $\eta'(\gamma\gamma)K$,
as well as the analog contributions from charmonium resonances 
($\eta_c$ and $\chi_c$ states).

The non-resonant contribution is negligible with respect to
the $K^*$ contribution everywhere in phase space.
Our evaluation of the amplitude given in \cite{Choudhury} 
confirms the small non-resonant branching fraction of order 10$^{-9}$
first reported by Hiller and Safir \cite{Hiller-Safir} 
in contradiction to the value given in \cite{Choudhury}.
Choudhury \emph{et al.} have recently acknowledged a numerical error in their
computations and published updated values \cite{Choudhuryrev} in accordance 
with \cite{Hiller-Safir}. 

The contributions from higher kaon resonances decaying to $K\gamma$
are difficult to assess with current experimental information.
Recent measurements of $B\rightarrow K_1(1270)\gamma$
and $K^*_2(1430)\gamma$ \cite{BKresexp} and corresponding radiative width
determinations for these resonances \cite{Cihangir-KTeV} indicate that the 
effective $K\gamma\gamma$ branching fractions from these higher resonances are
in the same range as for $K^*(892)$.
Since a number of other kaon resonances may contribute to
this final state, the overall $B\rightarrow K\gamma\gamma$ branching fraction 
due to kaon resonances could be an order of magnitude larger than 
our estimate for $K^*$ only, bringing it to a level that may
be accessible at currently running $B$ factories.
In view of the coarse experimental information available 
we leave these contributions to future investigations and assume here that 
their effects can be subtracted or isolated for the purpose of this study.

While the contributions from $\eta$ and $\eta'$ are sizable, giving effective 
branching fractions of up to 1.5 $\times$ 10$^{-6}$, they result in photons
of relatively low energy.
This will render the observation of interference effects with the 
$B\rightarrow K^*(K\gamma)\gamma$ amplitude experimentally difficult.
We will therefore focus on the more promising charmonium resonances
occurring at higher energies.
Table \ref{tab:Kccexp} summarizes the relevant experimental data for these
resonances.
Among the charmonium resonances, only $\eta_c$ and $\chi_{c0}$ are known both to
decay into two photons and to be produced in $B$ decays with an associated kaon,
so that we will restrict our analysis to these two resonances.

\begin{table}[tb]
    \begin{center}
        \caption{Branching fractions for the cascade decays 
         $B\rightarrow K (c\bar{c}) \rightarrow K\gamma\gamma$, where
         $(c\bar{c}) = \eta_c, \eta_c(\mathrm{2S}), \chi_{c0}, \chi_{c2}$, as far as they
         have been measured \cite{PDG,chiC-exp}.}
         \label{tab:Kccexp}
        \begin{tabular*}{\columnwidth}{@{\extracolsep{\fill}}lccc}
        \hline
        Resonance & 
        $\mathcal{B}_{(c\bar{c})\rightarrow \gamma\gamma}$ & 
        $\mathcal{B}_{B^0} \times \mathcal{B}_{(c\bar{c})}$ &
        $\mathcal{B}_{B^+} \times \mathcal{B}_{(c\bar{c})}$ \\
	& (10$^{-4}$) & (10$^{-7}$) & (10$^{-7}$) \\
        \hline
	$\eta_c$(2986)    & $4.3 \pm 1.5$   & $5.2 \pm 2.5$  & $3.9 \pm 1.8$ \\ 
	$\chi_{c0}$(3415) & $2.6 \pm 0.5$   & $<1.3$         & $0.8 \pm 0.2$ \\
        $\chi_{c2}$(3556) & $2.46 \pm 0.23$ & $<0.10$        & $<0.07$       \\
        $\eta_c(\mathrm{2S})$(3638) & --    & --             & --            \\
        \hline
       \end{tabular*}
    \end{center}
\end{table}

To model the amplitudes $\mathcal{M}_{\eta_c,\chi_{c0}}$ for the $B$ decays
to $\eta_c(\gamma\gamma)K$ and $\chi_{c0}(\gamma\gamma)K$ we use a general 
Breit-Wigner Ansatz along the lines described in Ref.~\cite{Hiller-Safir}.
Thus we neglect variations in the amplitudes beyond the Breit-Wigner form.
For $B\rightarrow \eta_c K$ we follow the factorization approach
employed in Ref.~\cite{Choudhury}.
The full amplitude for $B\rightarrow K \gamma\gamma$, including 
the three resonance contributions, is then given by 
\begin{equation} \label{eq:M}
 \mathcal{M}_{\mathrm{tot}} = 
 \mathcal{M}_{K^*} 
+ \xi_{\eta_{c}}\mathcal{M}_{\eta_{c}} 
+ \xi_{\chi_{c0}}\mathcal{M}_{\chi_{c0}},
\end{equation} 
where $\xi_{\eta_c,\chi_{c0}} = \pm 1$ denote unknown relative interference signs.
Note that in this simplified approach, the relative strong phases between the decay 
processes are assumed to be real.
While there are good reasons to question this assumption, we nevertheless choose to
study the relevant observables first in this approximation in order to investigate 
and illustrate the potential of the method \emph{in principle.} 
In Sec.~\ref{sec:strong} we will consider the case with arbitrary relative strong 
phases.

\section{Interference terms and asymmetries}

To study the role of the interference terms as photon polarization analyzers,
we generalize the SM amplitude for $B\rightarrow K^*(K\gamma)\gamma$
to include an amplitude for the emission of a right-handed photon from the $b$-quark.
Following Ref.~\cite{Hiller-Kagan}, we add a right-handed component to the operator 
$O_7$ from Eq.~\ref{eq:Heff}, \emph{i.e.} $C_7O_7\rightarrow C_7O_7+C'_7O'_7$,
with 
$O'_7= (e m_b)/(16\pi^2)\overline{s}_R \sigma_{\mu\nu} b_L F^{\mu\nu}$, 
describing the emission of a right-handed photon.
In this picture, the probability $f_R$ for the emission of a right-handed photon from the
$b$-quark is given by the corresponding Wilson coefficient,
$ f_R= |C'_7|^2/(|C_7|^2 + |C'_7|^2)$.
The naive SM estimate for this fraction is $f_R \approx 0.1\%$ based on 
$C'_7/C_7 \approx m_s/m_b$ from the chiral structure of 
the $W$-boson couplings to quarks \cite{Atwood-Gronau-Soni}.
A recent study including other operators that contribute to 
$b\rightarrow s\gamma_R$ finds that $f_R$ may be as large 
as 1\% within the SM \cite{GGLP}. 
In the following we assume the Wilson coefficients to be real, \emph{i.e.} 
we do not consider additional sources of CP violation beyond the SM.

\begin{figure}[t]
\center
\includegraphics*[width=8cm]{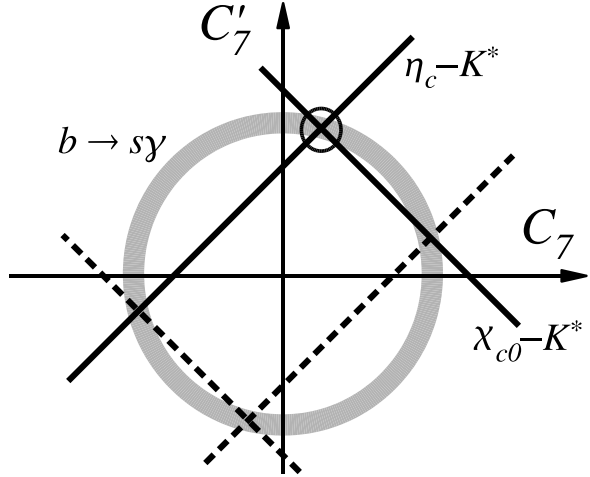}
\caption{Schematic illustration of the constraints in the $C_7$--$C'_7$
plane obtained from a spectrum measurement of $B\rightarrow K\gamma\gamma$
(under the assumption of negligibly small strong phases).
The gray circle depicts the region allowed from inclusive $b\rightarrow s\gamma$
measurements, the solid diagonal lines represent the solutions corresponding to
the $\eta_c$--$K^*$ and $\chi_{c0}$--$K^*$ interferences, with dashed lines
indicating mirror solutions in the case where the interference signs are unknown.} 
\label{fig:wheel}
\end{figure}

\begin{figure}[t]
\center
\includegraphics*[width=12cm]{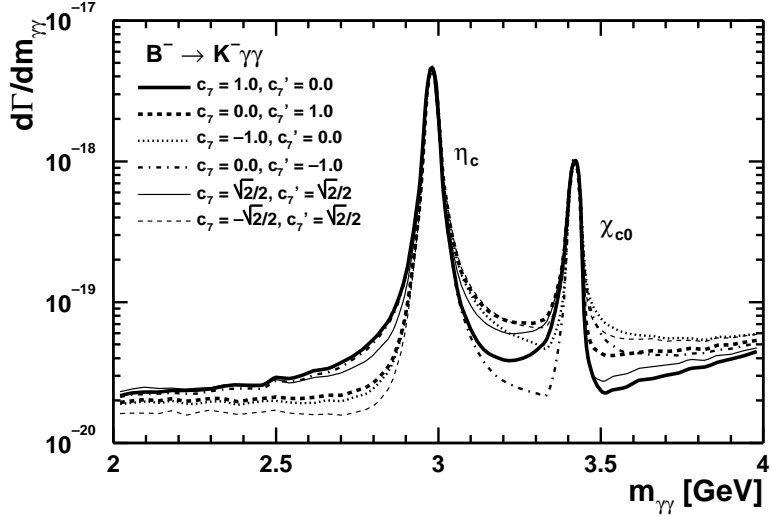}
\caption{Diphoton mass spectra for $B^-\rightarrow K^-\gamma\gamma$ 
for various values of the normalized Wilson
coefficients $c_7$ and $c'_7$.}
\label{fig:spectra}
\end{figure}

Taking account of the symmetry properties of the amplitude (\ref{eq:SMamp})
it is straightforward to incorporate the emission of a right-handed 
photon by adding a parity-inverted term proportional to $C'_7$.
Evaluation of the full amplitude then shows explicitly 
that the $\eta_c$--$K^*$ interference term is proportional to $(C_7-C'_7)$ 
while the $\chi_{c0}$--$K^*$ interference term is proportional to $(C_7+C'_7)$.
These interference terms are accessible to experiment:
they manifest themselves as enhancements or reductions in the diphoton mass 
spectrum of $B\rightarrow K\gamma\gamma$ decays near the resonance peaks, 
depending on the signs involved.
The Wilson coefficients $C_7$ and $C'_7$ may thus be 
cleanly extracted from the observed diphoton mass spectrum, if the signs of 
the interference terms are known (and relative strong phases are negligible,
see Sec.~\ref{sec:strong}).
Unfortunately, neither of the two interference signs is known 
model-independently today, such that, even under the assumption of negligibly
small strong phases, only values for $|C_7-C'_7|$ and $|C_7+C'_7|$ could be 
derived from a measured spectrum, leading to a four-fold ambiguity in the 
solution for $(C_7,C'_7)$, see Fig.~\ref{fig:wheel}.
In spite of the four-fold ambiguity, a measurement of these interference terms
may still represent a valuable test of the SM.
Recall that the overall normalization $|C_7|^2 + |C'_7|^2$ is given by the
inclusive $b\rightarrow s\gamma$ rate and hence already known from experiment.
In Fig.~\ref{fig:spectra} we show diphoton mass spectra for various values of 
$c_7\mbox{}^{(}\mbox{}'\mbox{}^{)} = 
 C_7\mbox{}^{(}\mbox{}'\mbox{}^{)} /\sqrt{C_7^2 + C_7'^2}$ 
for $B^-$ decay, with positive interference signs assumed throughout.

\begin{figure}[t]
\center
\includegraphics*[width=12cm]{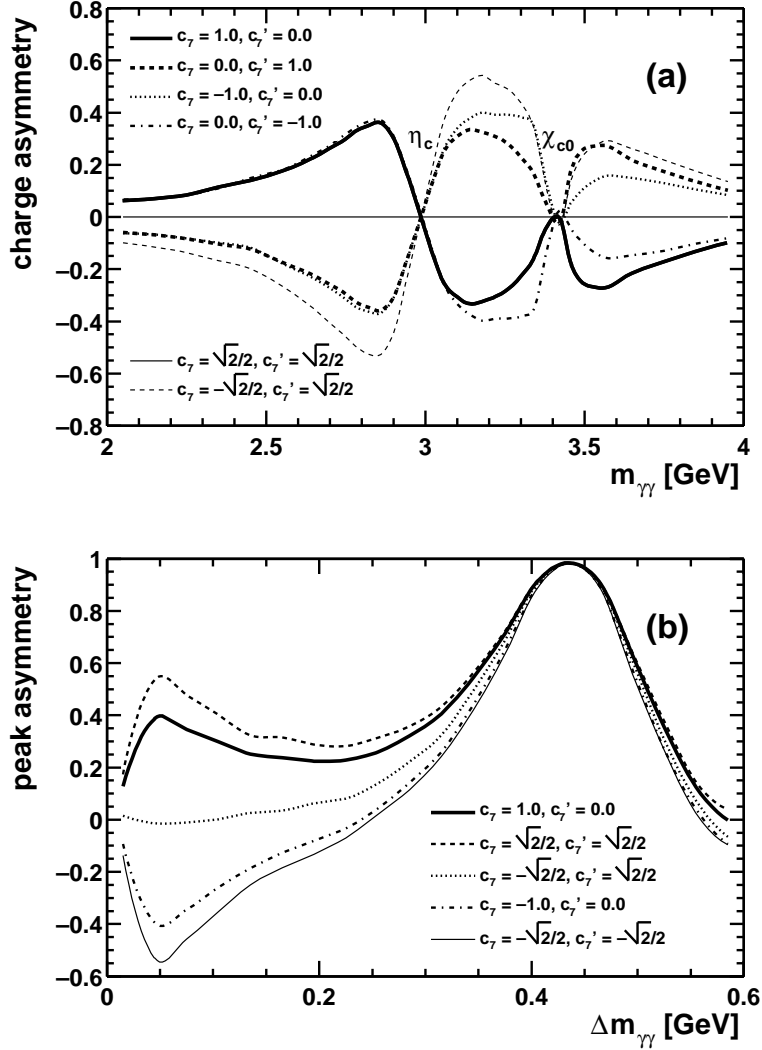}
\caption{(a) Charge asymmetry $A_C(m_{\gamma\gamma})$ 
and (b) peak asymmetry $A_{\chi_{c0}}(\Delta m_{\gamma\gamma})$
for various values of $c_7$ and $c'_7$. 
All interference terms are assumed to have positive sign.} 
\label{fig:asym}
\end{figure}

Experimentally, the interference terms are most readily isolated by means of
asymmetries.
An observable that is particularly convenient to extract the $\eta_c$ interference 
is the charge asymmetry $A_C$, defined as
\begin{equation}
  A_C(m_{\gamma\gamma}) = 
    \frac{d\Gamma^-/dm_{\gamma\gamma} - 
          d\Gamma^+/dm_{\gamma\gamma}}
         {d\Gamma^-/dm_{\gamma\gamma} +
          d\Gamma^+/dm_{\gamma\gamma}} ,
\end{equation}
with $\Gamma^{\pm} = \Gamma(B^{\pm}\rightarrow K^{\pm}\gamma\gamma)$.
An analog asymmetry may be defined for neutral $B$ decays, where 
experimental difficulties arise from flavor tagging, compensated in part 
by the larger statistics available.
Here we only consider the charged decay.
In Fig.~\ref{fig:asym} (a) we show the expected $A_C$ 
for various combinations of $c_7$ and $c'_7$.
It exhibits the typical shape of a Breit-Wigner interference around the position
of the $\eta_c$ resonance, with a distortion at higher energies due to the presence 
of the $\chi_{c0}$ resonance, which forces the charge asymmetry to zero in its 
vicinity.
The value of the maximum asymmetry below the $\eta_c$ peak 
is a direct measure of $(c_7-c'_7)$, we find
$A_C^\text{max} \approx (0.37\pm 0.02)\times (c_7-c'_7)$ for positive interference sign.
The error is dominated by the uncertainty in the $\eta_c$ branching fraction, 
see Table~\ref{tab:Kccexp}.

The $\chi_{c0}$ interference, being CP-even, cannot be extracted in the same
manner.
Instead we define a charge-averaged peak asymmetry around the $\chi_{c0}$,
\begin{equation}
  A_{\chi_{c0}} (\Delta m_{\gamma\gamma}) = 
    \frac{d\overline{\Gamma}(m^-)/dm_{\gamma\gamma} - 
          d\overline{\Gamma}(m^+)/dm_{\gamma\gamma}}
         {d\overline{\Gamma}(m^-)/dm_{\gamma\gamma} + 
          d\overline{\Gamma}(m^+)/dm_{\gamma\gamma}},
\end{equation}
where $m^{\pm} = m_{\chi_{c0}}\pm\Delta m_{\gamma\gamma}$, and
$\overline{\Gamma} = (\Gamma^++\Gamma^-)/2$.
The expected peak asymmetry is shown in Fig.~\ref{fig:asym} (b), again for
various combinations of $c_7$ and $c'_7$.
It is dominated by the sought-after interference effect, since the 
distribution of $B\rightarrow K^*(K\gamma)\gamma$ events is rather flat
in that region.
For values of $\Delta m_{\gamma\gamma}$ well below 
$m_{\chi_{c0}}-m_{\eta_c}$ (at which point $A_{\chi_{c0}} = 1$ due to the $\eta_c$
peak) we find $A_{\chi_{c0}}^\text{max} \approx (0.40\pm 0.01)\times (c_7+c'_7)$
for positive interference sign. 
In this case the error mainly originates from the uncertainty in the $\chi_{c0}$ 
branching fraction (Table~\ref{tab:Kccexp}).

\section{Uncertainty from strong phases}
\label{sec:strong}

In our simplified approach within the factorization approximation we have neglected
the effect of relative strong phases between the $B\rightarrow K^* \gamma$ and 
$B\rightarrow \eta_c (\chi_{c0}) K$ decays.
Recent evaluations for $B\rightarrow D\pi$ \cite{BDpi} and $B\rightarrow\pi\pi$ 
\cite{Bpipi}, however, indicate sizable strong phases, 
thus casting into doubt our initial assumption.

In the presence of strong-phases the coefficients $\xi_{\eta_c,\chi_{c0}}$ in 
Eq.~\ref{eq:M} simply become $\exp(i\phi_{\eta_c,\chi_{c0}})$ where 
$\phi_{\eta_c,\chi_{c0}}$ denote the relative strong phases between the 
$B\rightarrow K^* \gamma$ and $B\rightarrow \eta_c (\chi_{c0}) K$ amplitudes.
The corresponding interference terms appearing in the $K\gamma\gamma$ spectrum
will no longer only depend on $(C_7-C'_7)$ and $(C_7+C'_7)$, but rather on
$\cos\phi_{\eta_c}(C_7-C'_7)$ and $\cos\phi_{\chi_{c0}}(C_7+C'_7)$. 
Thus the extraction of useful information on $C_7$ and $C'_7$ entirely hinges
on the knowledge of the relative strong phases $\phi_{\eta_c}$ and $\phi_{\chi_{c0}}$.
This severely limits the applicability of the method for the time being.

Conversely, we may of course note that once the photon polarization is known from 
one of the other proposed methods, a measurement of the above defined asymmetries 
may serve to improve our understanding of the strong phases at play.

\section{Experimental considerations and conclusion}

Apart from the strong phase problem, the principal experimental limitation for such 
a measurement will be the required statistics of $B$ decays.
To arrive at a rough estimate of the required order of magnitude of
$B$ mesons, we note that some $10^3$ clean 
$B\rightarrow K^*(K\gamma)\gamma$ decays would be necessary
for a measurement distinguishing between the case of maximum asymmetry
from that of zero asymmetry.
Factoring in branching fractions and typical reconstruction efficiencies for
radiative decays at $e^+e^-$ $B$ factories ($\approx$10\%) and hadron 
colliders ($\approx$0.1\%) we find that several 10$^{10}$ (10$^{12}$) 
neutral or charged $B$ mesons would be needed in the case of 
an $e^+e^-$ (hadron) collider. 
These numbers are compatible with expected annual production rates at
future facilities being proposed \cite{SuperB} or built \cite{LHCbReoptTDR}.
Of course, many experimental issues remain to be addressed within the context 
of a specific experimental setup and as more knowledge on the amplitudes 
involved in $B\rightarrow K\gamma\gamma$ decay becomes available.

Our main conclusion is that contributions from kaon resonances dominate
the $B\rightarrow K\gamma\gamma$ yield throughout most of the phase space
and thus render the non-resonant $b\rightarrow s\gamma\gamma$ amplitude
inaccessible to experiment in this final state.

Furthermore, we have investigated the possibility to utilize 
resonance interferences in the $K\gamma\gamma$ final state to 
probe the photon polarization in the $b\rightarrow s\gamma$ transition,
which may reveal contributions from new physics beyond the SM.
While possible in principle, the method suffers in practice from theoretical
uncertainties related to the unknown strong phases present in the decays
and experimentally from the formidable requirement on the statistics of $B$ 
meson decays.
But in the event that the relevant strong phases can be obtained from
elsewhere and the required number of $B$ decays can be collected, the method has 
the advantage of yielding direct information on the Wilson coefficients
$C_7$ and $C'_7$.

\section*{Acknowledgments}

We are indebted to Gudrun Hiller for very helpful suggestions and comments.
We also thank Namit Mahajan for fruitful exchanges.
For some of our calculations we have used the EvtGen software package
\cite{EvtGen}.
This work was supported by the Swiss National Science Foundation under
grant Nr.~620-066162.

\end{document}